\input{aipcheck}

\documentclass[
    final            
  ]
  {aipproc}

\layoutstyle{6x9}

\begin{document}

\title[Testing the \textsc{SAtlas} Code with Interferometric Observations]{Modeling Stellar Atmospheres with a Spherically Symmetric Version of the \textsc{Atlas} Code: Testing the Code by Comparisons to Interferometric Observations and \textsc{Phoenix} Models}


\author{Hilding R. Neilson}{
  address={Department of Astronomy \& Astrophysics, University of Toronto, Toronto, ON, Canada \\
  {\tt neilson@astro.utoronto.ca}}
}

\author{John B. Lester}{
}
\begin{abstract}
One of the current opportunities for stellar atmospheric modeling is the interpretation of optical interferometric data of stars. Starting from the robust, open source \textsc{Atlas} atmospheric code \citep{Kurucz1970}, we have developed a spherically symmetric code, \textsc{SAtlas}, as a new option for modeling stellar atmospheres of low gravity stars. The \textsc{SAtlas} code is tested against both interferometric observations of M giants by Wittkowski and collaborators, and spherically symmetric M giant NextGen models from the \textsc{Phoenix} code. The \textsc{SAtlas} models predict interferometric visibilities that agree with the observed visibilities and with predicted visibilities, and the  \textsc{SAtlas} atmospheric structures also agree with those from spherical  \textsc{Phoenix} models, with just small differences in temperature and pressure at large depths in the atmospheres.
\end{abstract}
\maketitle

\section{Introduction}
Cool giant stars have low gravity, and are very luminous.  They form a large fraction of the stellar population and it is important to understand these stars as a part of understanding stellar structure and evolution.

Optical interferometric observations are a powerful tool for understanding these stars and the ongoing development of interferometry is providing more precise details about the structure of stellar atmospheres.  These observations also precisely measure limb--darkened angular diameters of stars and the center-to-limb variation of the intensity.  This was done using K-band  Very Large Telescope Interferometer (VLTI/VINCI) observations of $\psi$ Phe, $\gamma$ Sge, and $\alpha$ Cet to test spherically symmetric \textsc{Phoenix} stellar atmosphere models assuming local thermodynamic equilibrium (LTE) and plane parallel \textsc{Atlas} models \citep{Wittkowski2004, Wittkowski2006a, Wittkowski2006b}.  

In this preliminary work we model the K-band VLTI/VINCI observations using a new version of the LTE stellar atmospheres code \textsc{Atlas} that is spherically symmetric \citep{Lester2008}.  We also compare the structure of the models with effective temperature and gravities of these stars with \textsc{Phoenix} models.  The purpose of this analysis is to test how robust the SAtlas code is at low gravity by comparing models to observations and to \textsc{Phoenix} models, which is a benchmark for modeling stellar atmospheres.  

\section{Fitting Models to Interferometric Data}
We use the results of \citep{Wittkowski2004, Wittkowski2006a, Wittkowski2006b} to determine the range of luminosity, mass, and radius for each star, shown in Table 1, and calculate grids of spherically symmetric model stellar atmospheres .

\begin{table}
\caption{Input Parameters for the 3--D Grid of Model Stellar Atmospheres for Fitting Each Star.}
\begin{tabular}{lccccc}
\hline
Star & $L_{\rm{Min}}(L_\odot)$  & $L_{\rm{Max}}(L_\odot)$  & $\Delta L(L_\odot)$ & $M_{\rm{Min}}(M_\odot)$  & $M_{\rm{Max}}(M_\odot)$  \\
\hline
$\psi$ Phe & $630$ & $1580$ & $50$ &  $0.6$ & $1.6$ \\
$\gamma$ Sge & $400$&$700$&$50$&$1.0$&$1.9$ \\
$\alpha$ Cet & $100$&$2000$&$ 50$&$ 1.5$&$ 3.0$\\
\hline
 & $\Delta M(M_\odot)$ & $R_{\rm{Min}}(R_\odot)$  & $R_{\rm{Max}}(R_\odot)$  & $\Delta R(R_\odot)$ \\
 \hline
$\psi$ Phe &  $0.2$ & $60$ & $120$& $20$ \\
$\gamma$ Sge&$ 0.2$&$60$&$120$&$10$ \\
$\alpha$ Cet&$ 0.1$&$ 60$&$ 100$&$ 10$ \\
\hline
\end{tabular}
\end{table}
\begin{figure}
  \includegraphics[height=.24\textheight]{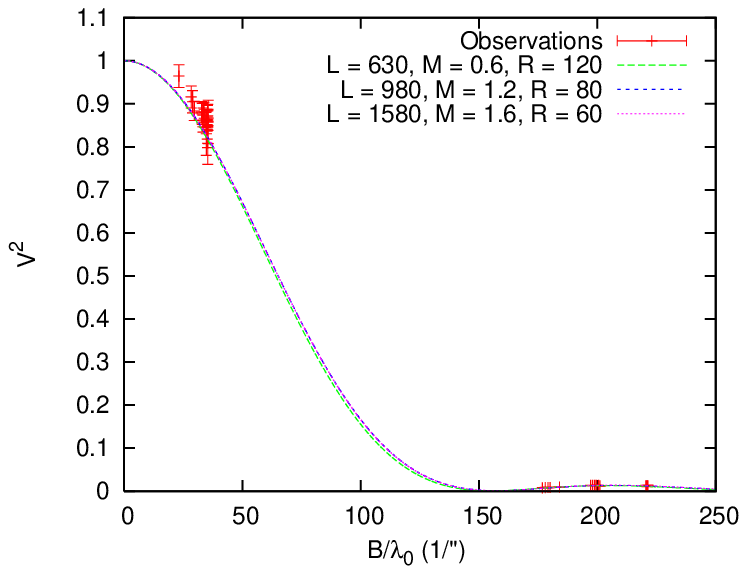}\includegraphics[height=.24\textheight]{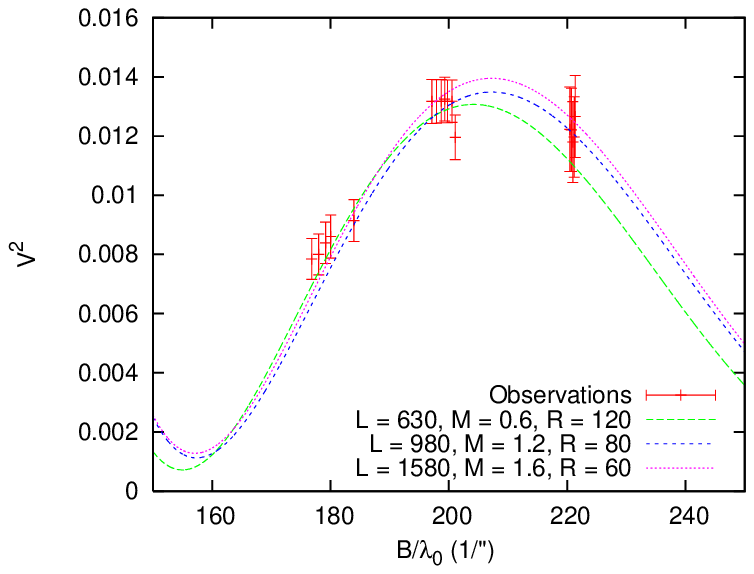}
  \caption{Comparison of the predicted visibilities for three of the model stellar atmospheres with the observed data of $\psi$ Phe, (Right) with a close--up of the second lobe.}
\end{figure}
\begin{figure}
  \includegraphics[height=.24\textheight]{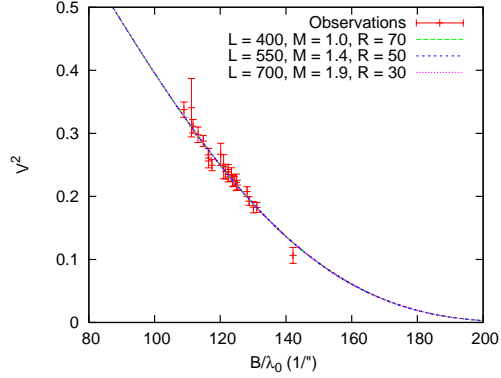}
  \caption{Comparison of the predicted visibilities for three of the model stellar atmospheres with the observed data of $\gamma$ Sge.}
\end{figure}
\begin{figure}
  \includegraphics[height=.24\textheight]{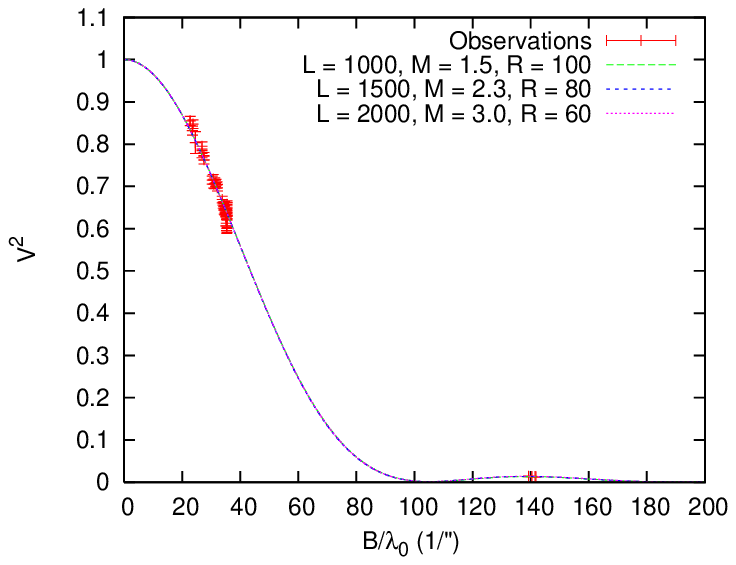}\includegraphics[height=.24\textheight]{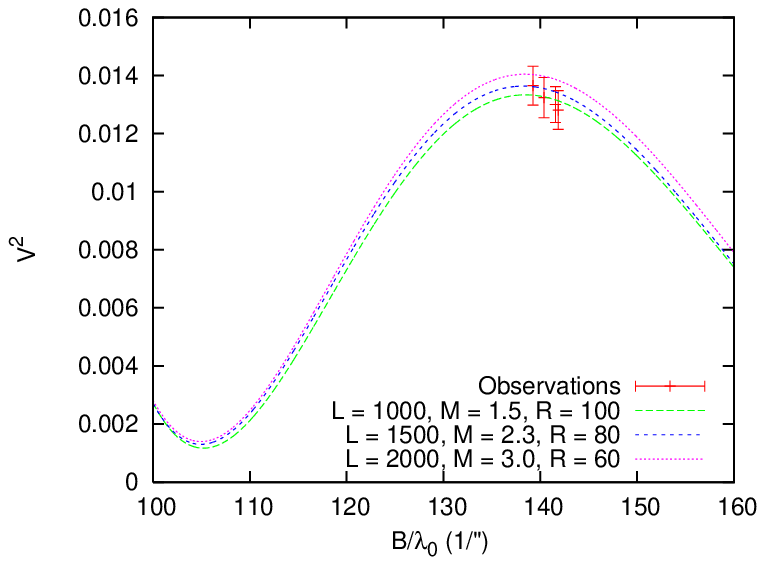}
  \caption{Comparison of the predicted visibilities for three of the model stellar atmospheres with the observed data of $\alpha$ Cet, (Right) with a close--up of the second lobe.}
\end{figure}

To model the visibilities, we produce a grid of 480 models for $\psi$ Phe, 350 models for $\gamma$ Sge, and 1680 models for $\alpha$ Cet (there are more models for this case to test the fit over a much larger range).  We find the best-fit limb-darkened angular diameter for each star using $\chi^2$ fitting and produce the Rossland angular diameter $\theta_{\rm{Ross}}$, the layer with an Rossland optical depth of unity.  The best-fit value for $\psi$ Phe is $\theta_{\rm{Ross}} = 8.13 \pm 0.1 \rm{mas}$ with $\chi^2 = 1.67$, for $\gamma$ Sge is $\theta_{\rm{Ross}} = 6.02 \pm 0.01 \rm{mas}$ with $\chi^2 = 0.64$, and for $\alpha$ Cet is $\theta_{\rm{Ross}} = 12.1 \pm 0.05 \rm{mas}$ with  $\chi^2 = 0.98$.  This is consistent with earlier results found using \textsc{Phoenix} and plane-parallel \textsc{Atlas} models.  For the case of $\psi$ Phe, the $\chi^2$ value is smaller than that predicted using the \textsc{Phoenix} code, though we predict the same value of $\theta_{\rm{Ross}}$.   The \textsc{Phoenix} and \textsc{SAtlas} fits agree within the observational uncertainty.  We also predict a smaller value of $\theta_{\rm{Ross}}$  for $\alpha$ Cet, with a difference of $ 0.1 \rm{mas}$.  The differences may be related to the fact that we are using a larger number of models to fit the observations and  to differences between \textsc{Phoenix} and spherical \textsc{Atlas} models. This may be an example of interferometric observations highlighting small differences in computed intensity distributions between stellar atmosphere codes.

\section{Comparison With \textsc{Phoenix} Models}
The comparison of the spherical \textsc{Atlas} models to interferometric observations show that the new spherically symmetric version of the \textsc{Atlas} code is a robust tool for studying stellar atmosphere; however it is important to compare this to similar \textsc{Phoenix} models.  Here we compare NextGen models \citep{Hauschildt1999} with $T_{\rm{eff}} = 3800$, $3500$ $K$ and $\log g (cm/s^2) = 0.5$, $1$ and mass $M = 2.5$ and $5.0 M_\odot$. 

\begin{figure}\label{f5}
  \includegraphics[height=.24\textheight]{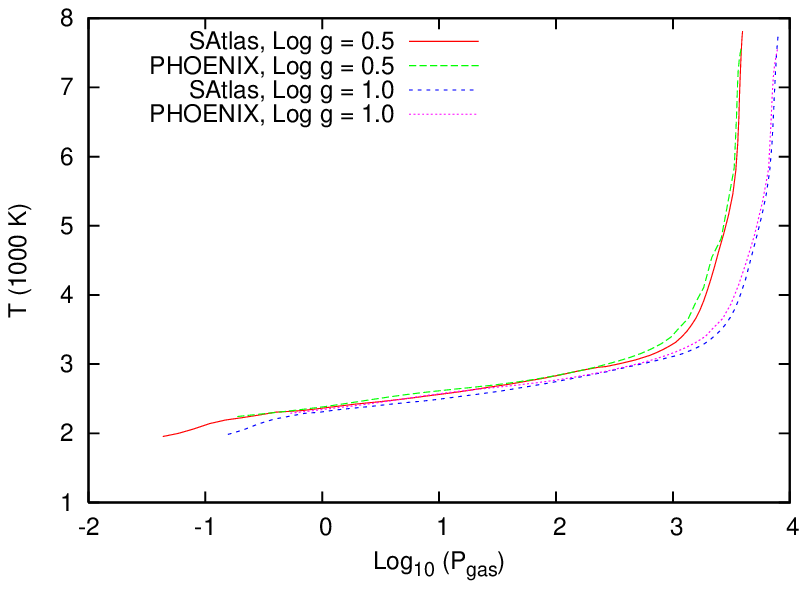}\includegraphics[height=.24\textheight]{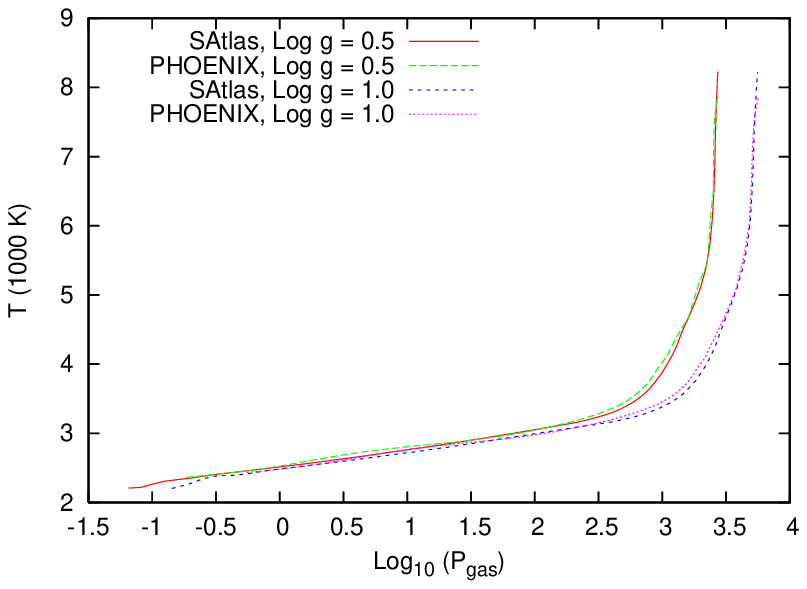}
  \caption{Comparison of the temperature structure of model stellar atmospheres computed by the spherical \textsc{Atlas} code and \textsc{Phoenix} NextGen models for two gravities $\log g = 0.5$, $1.0$. (Left) The models have an effective temperature of $3400$ $K$ and mass of $5M_\odot$. (Right) The models have an effective temperature of $3800$ $K$ and mass of $5 M_\odot$.}
\end{figure}

By comparing the temperature structure as a function of the gas pressure, we can test if there exist differences in the properties of the computed stellar atmospheres that would also produce differences in the center-to-limb intensity structure. It is shown in Figure 4 that models produced by the \textsc{SAtlas} code are consistent with spherically symmetric, LTE \textsc{Phoenix} models.  There are only small differences between the temperature structure in this effective temperature and gravity range, most likely due to differences in the codes.  This comparison shows that the Spherical \textsc{Atlas} code is robust and models low gravity stellar atmospheres in LTE as well as \textsc{Phoenix}. 

\section{Conclusions}
The goal of this work was to test new, updated version of the \textsc{Atlas} code by comparing stellar atmosphere models with interferometric observations and comparing these results with results using the \textsc{Phoenix} code.  This work also directly compares the structures of cool, low gravity stellar atmosphere models computed using the spherical \textsc{Atlas} code with NextGen \textsc{Phoenix} models.
  
We find that the spherical \textsc{Atlas} models fit the interferometric observations as well as \textsc{Phoenix} models, and for case of $\psi$ Phe, the spherical \textsc{Atlas} models fit a little better. By directly comparing our models to NextGen models, we see negligible differences in the temperature structure.  It is shown that the spherical \textsc{Atlas} code is a robust tool for stellar atmosphere studies.  Please contact us if you would like a copy of the spherical \textsc{Atlas} code.

\bibliographystyle{aipproc}
\bibliography{cs15}

\end{document}